# Trapping of classical particles by an electromagnetic radiation amplifying with time

Azad Ch. Izmailov

Institute of Physics, Azerbaijan National Academy of Sciences, Javid av. 33, Baku, Az-1143, AZERBAIJAN

*e-mail*: azizm57@rambler.ru

**ABSTRACT.** We analyze new possible applications of the trapping mechanism of sufficiently slow-speed particles by an electromagnetic potential well deepening with time (up to a certain limit) which was recently established by author from basic relations of classical mechanics. It is assumed that given particles are under conditions of the high vacuum and forces acting on these particles are not dissipative. Such wells may be created by means of an electromagnetic field (in particular radiation) with a fixed spatial distribution but with a nondecreasing strength in time. Trapping and localization of particles in such electromagnetic traps are analyzed on example of typical laser beams. Obtained results may be used in high resolution spectroscopy of various particles (including also atoms and molecules in definite cases).

Electromagnetic traps for charged and neutral particles without material walls allow to localize and observe these particles during a comparatively long period of time thereby creating conditions for detailed research of their properties [1]. In particular, such traps of particles in the high vacuum open new possibilities for contactless measurements of forces acting on given particles with extremely high accuracy and allow micromanipulations of these particles [2]. Even more important is the development of effective methods of trapping and localization of sufficiently slow-speed atoms, molecules and ions for ultrahigh resolution spectroscopy [3] and for creation of more precise standards of time and frequency [4].

Recently the new trapping mechanism of sufficiently slow-speed particles by an electromagnetic potential well deepening with time (up to a certain limit) was established by author from basic relations of classical mechanics [5]. Such potential wells may be induced by an intensifying with time electromagnetic field having a fixed spatial distribution. It is assumed that considered particles are under conditions of the high vacuum and forces acting on given particles are not dissipative (that is motion of these particles occurs without friction). Depending on whether particles have electric or magnetic moment, it is possible to use the controllable electric or magnetic field or nonresonance laser radiation for their trapping by such a method. Earlier author has investigated the considered trapping mechanism in the comparatively simple model of the one-dimensional rectangular potential well deepening with time [5]. In the present work we will analyze features of trapping and three-dimensional localization of classical particles by real characteristic laser beams with amplifying intensity. Obtained results may be used for motion control and high-resolution spectroscopy of various particles including also atoms and molecules in definite cases.

According to my work [5], we consider a point particle with the mass $m$ freely moving in a three-dimensional space before its entering to the region $V$ of the potential well $U(\boldsymbol{R},t)$, which explicitly depends not only on the coordinate $\boldsymbol{R}$ but also on time $t$. The total energy of such a particle with the non-relativistic velocity $\boldsymbol{v}$ is described by the known formula [6]:

$$E(\boldsymbol{R},\boldsymbol{v},t) = 0.5mv^2 + U(\boldsymbol{R},t). \qquad (1)$$

Further we will consider the potential energy $U(\boldsymbol{R},t)$ of the following form:

$$U(\boldsymbol{R},t) = s(\boldsymbol{R}) * \varphi(t), \qquad (2)$$

where the coordinate function $\sigma(\boldsymbol{r}) \leq 0$ in the region $V$, and $\varphi(t) \geq 0$ is the nondecreasing function of time $t$. Such a potential (2) may be created for particles having electric or magnetic moment by a controllable electromagnetic field with the growing strength (up to a certain time moment) but with a fixed spatial distribution [7]. We have the following motion equation of the particle in case of the potential energy (2):

$$m\frac{d^2\boldsymbol{R}}{dt^2} = -\varphi(t)\frac{ds(\boldsymbol{R})}{d\boldsymbol{R}}. \qquad (3)$$

From relations (1)-(3) we directly receive the formula for the time derivative of the total energy $E(\boldsymbol{R},\boldsymbol{v},t)$ of the particle:

$$\frac{dE}{dt} = s(\boldsymbol{R})\frac{d\varphi(t)}{dt} \leq 0. \qquad (4)$$

According to inequality (4), increase of the function $\varphi(t)$ with time $t$ leads to decrease of the total energy $E(\boldsymbol{R},\boldsymbol{v},t)$ (1) of the particle in the region $V$ of the potential well, where the coordinate function $s(\boldsymbol{R}) \leq 0$. We also see from the formula (1) that the particle can not go beyond the potential well and reach the region with $U(\boldsymbol{R},t) = 0$, when its total energy $E$ will be negative. It is important to note, that under such a condition $E < 0$, the considered classical particle will be localized in the region $V$ of the potential well even after output of the nondecreasing time function $\varphi(t)$ on a constant value. Detailed research of dynamics of particles may be carried out on the basis of relations (1)-(4) for electromagnetic traps with definite spatial configurations.

Further we will analyze trapping and three-dimensional localization of such classical particles by a nonhomogeneous electromagnetic radiation which is amplified with time (up to a certain limit). It is assumed that the force of the light pressure from this radiation is negligible in comparison with its light induced gradient force affected on given particles. Similar situations are possible for transparent particles in a spectral range of their irradiation. We will consider not too strong radiation, when an induced electric dipole moment of a particle is proportional to the light field strength and the potential energy of such a particle, respectively, is proportional to the square of this strength [2,7].

Let us analyze the possible (for practical realization) case of the amplifying with time standing light wave (along the axis z) whose intensity has the transversal Gaussian distribution. Such a radiation creates the potential well of the type (2) with the following coordinate function $s(\mathbf{R})$ for particles with light induced dipole moments [2,3]:

$$s(\mathbf{R}) = -J_0 * exp(-r^2/r_0^2) * sin^2(kz), \qquad (5)$$

where $r = \sqrt{x^2 + y^2}$ is the distance from the central axis of the light beam (with the characteristic radius $r_0$), $k$ is the wave number, and $J_0 > 0$ is the value with the dimension of energy, which is determined by a polarizability of a particle. For example, we will consider the following time dependence $\varphi(t)$ (2) for the radiation intensity:

$$\varphi(t) = (t/T) * \eta(T - t) + \eta(t - T), \qquad (t \geq 0), \qquad (6)$$

where $\eta(q)$ is the step function ($\eta(q) = 1$ if $q \geq 0$ и $\eta(q) = 0$ when $q < 0$). The function $\varphi(t)$ (6) linearly increases from 0 to 1 in the interval $0 \leq t \leq T$ and is equal to unit, when $t > T$.

Fig.1,a presents numerically calculated (on the basis of the motion equation (3)), temporary dependences of the distance $r(t)$ from the axis of the light beam and longitudinal coordinate $z(t)$ of the particle, which approaches from the outside to the given beam at starting conditions specified in the moment $t_0=0$. We see, that during increasing of the radiation intensity with time $t$ (6), trapping and three-dimensional localization of the considered particle occurs in the region of the beam which, according to the function $s(\mathbf{R})$ (5), creates the spatially periodic potential along the axis $z$. In this case the particle carries out vibration transversal motions $r(t)$ in limits determined by the characteristic radius $r_0$ of the light beam (fig.1,a) and also undergoes comparatively fast nondamped oscillations $z(t)$ in the longitudinal direction in the limit of the half wavelength $\lambda = 2\pi/k \ll r_0$ of the radiation (fig.2). The given particle remains localized in such a potential well even if $(t - t_0) > T$ (fig.1,a), that is after the output of the radiation intensity on the constant value according to formula (6). One can see from comparison of fig.1,a and dependence 1 in fig.3 that trapping of this particle occurs when its total energy $E(t)$ (1) decreases up to negative values because of the inequality (4).

Fig.1, b presents dynamics of a particle with the same initial values of velocity and coordinates as in fig.1, a but specified in a moment $t_0 > T$. Then, according to the dependence (6), the given particle flies through the stationary light beam, whose intensity is equal to the maximum value for the considered case of fig.1, *a*. We see that this particle is not captured by the light beam and moves away from it after the primary rapprochement (fig.1,*b*). The total energy $E(t)$ (1) of the given particle is constant during the whole its movement (dependence 2 in fig.3).

Author has carried out numerical calculations also for a number of other potential wells with cylindrical and spherical symmetries, which are described by the general formula (2). These calculations confirmed following qualitative results (*a*), (*b*) and (*c*) of the present work:

(*a*) Even a highly shallow but increasing with time potential well may continuously capture sufficiently slow-speed particles flying through it.

(*b*) Such trapped particles will remain in this potential well even after going out of a nondecreasing strength of the corresponding electromagnetic field on stationary values. However such a stationary trap already will not capture new particles.

(*c*) Since considered electromagnetic traps are based on non-dissipative forces, then particles, captured in given traps, carry out non-damped oscillation motions in limits of corresponding potential wells.

We have considered structureless classical particles in potential wells of the certain type (2). In practice it is possible, for example, for a collection of noninteracting (with each other) microparticles, which fly without friction under conditions of the ultrahigh vacuum at action of the controllable electric (magnetic) field or laser radiation with fixed configuration.

For analysis of possible capture of atoms and molecules in such traps, consideration of their quantum structure is necessary. At the same time, results obtained in this work may be applied also for such atomic objects in definite cases. Thus, for example, it is possible creation of traps for atoms and molecules by a nonhomogeneous laser radiation with frequencies essentially detuned from resonances with atomic (molecular) transitions [2,8,9]. Then the gradient force acts on atoms (molecules), which are in the ground quantum state, in the direction to the point of minimum of the light induced potential well. In particular, such wells may be induced also by laser beams considered in the present work. However, as was shown above, even highly slow-speed microparticles, flying from outside through such stationary beams, will not be captured in corresponding potential wells. At the same time, proposed intensification of the laser radiation (during a certain time interval) will lead to large increase of a number of particles, captured in given traps. This will allow essentially extend possibilities of ultra-high resolution spectroscopy of various microparticles (including atoms and molecules).

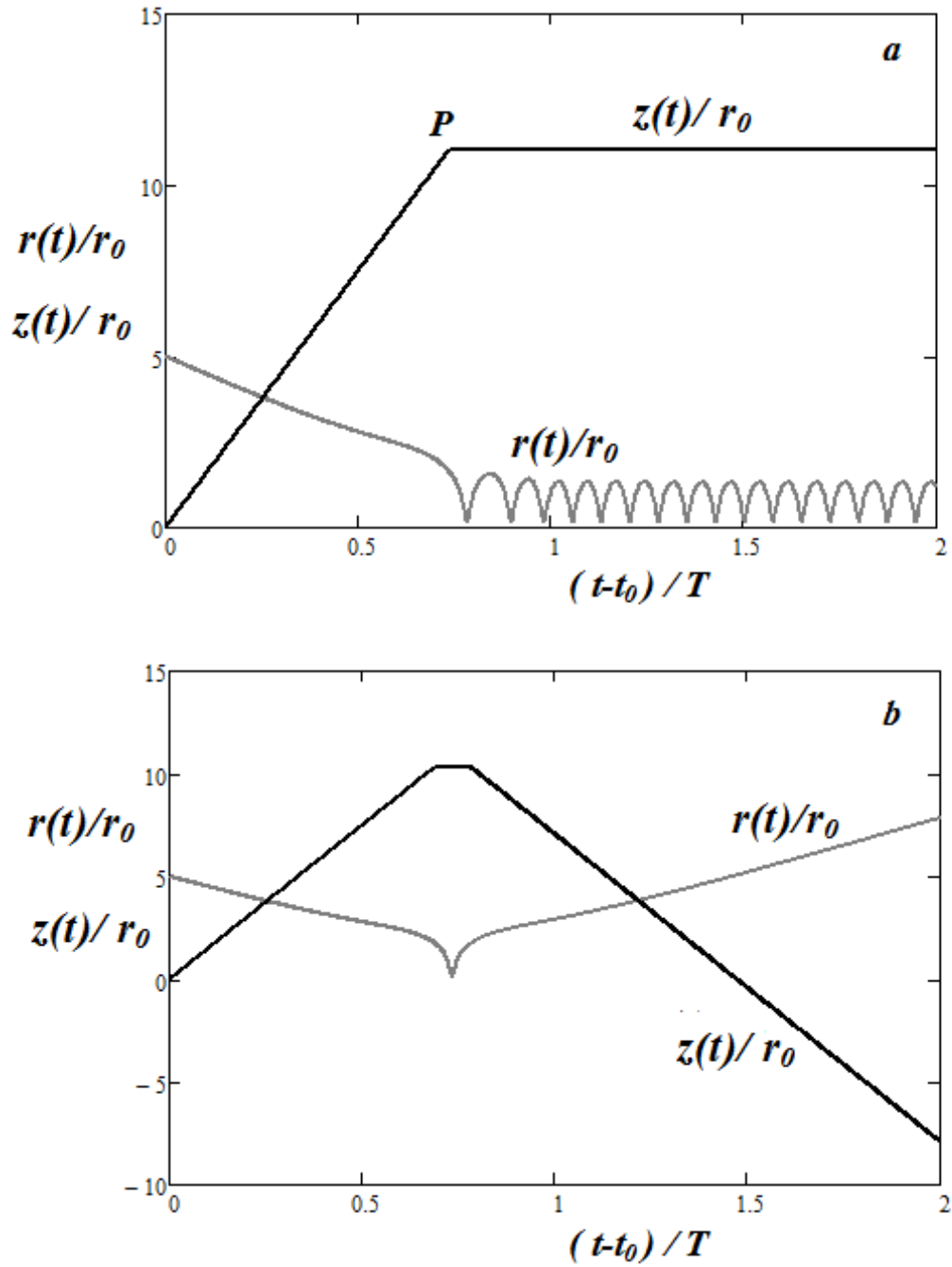

**Fig.1.** Dependences of particle coordinates $r(t) = \sqrt{x(t)^2 + y(t)^2}$ and $z(t)$ on time $t \geq t_0$ at particle initial coordinates $x(t_0)=0$, $y(t_0) = 5r_0$, $z(t_0)=0$ and its initial velocity components $v_x(t_0) = -2.5(r_0/T)$, $v_y(t_0) = -5(r_0/T)$, $v_z(t_0) = 15(r_0/T)$, given in moments $t_0 = 0$ (*a*) and $t_0 > T$ (*b*), when $J_0 = 5000m(r_0/T)^2$ and $kr_0=1000$.

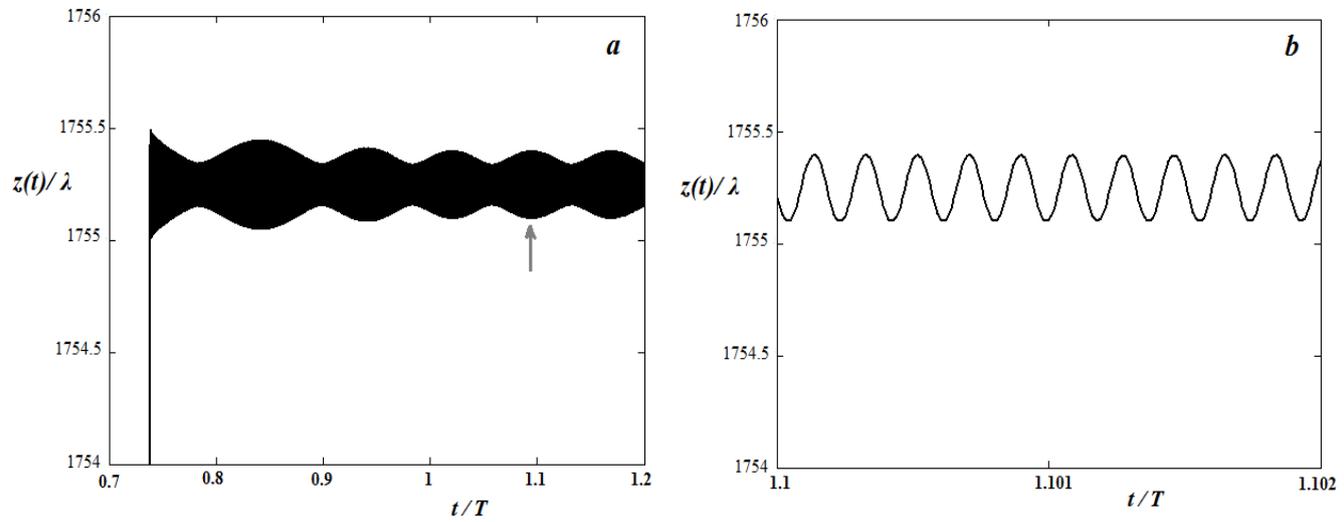

**Fig.2.** Particle coordinate $z(t)$ (in units of the radiation wavelength $\lambda = 2\pi/k$) versus time $t$ in enlarged scales in the neighborhood of the point $P$ from fig.1,a. Fig. 2,b presents the dependence $z(t)$ in the narrow region indicated by the arrow in fig. 2,a.

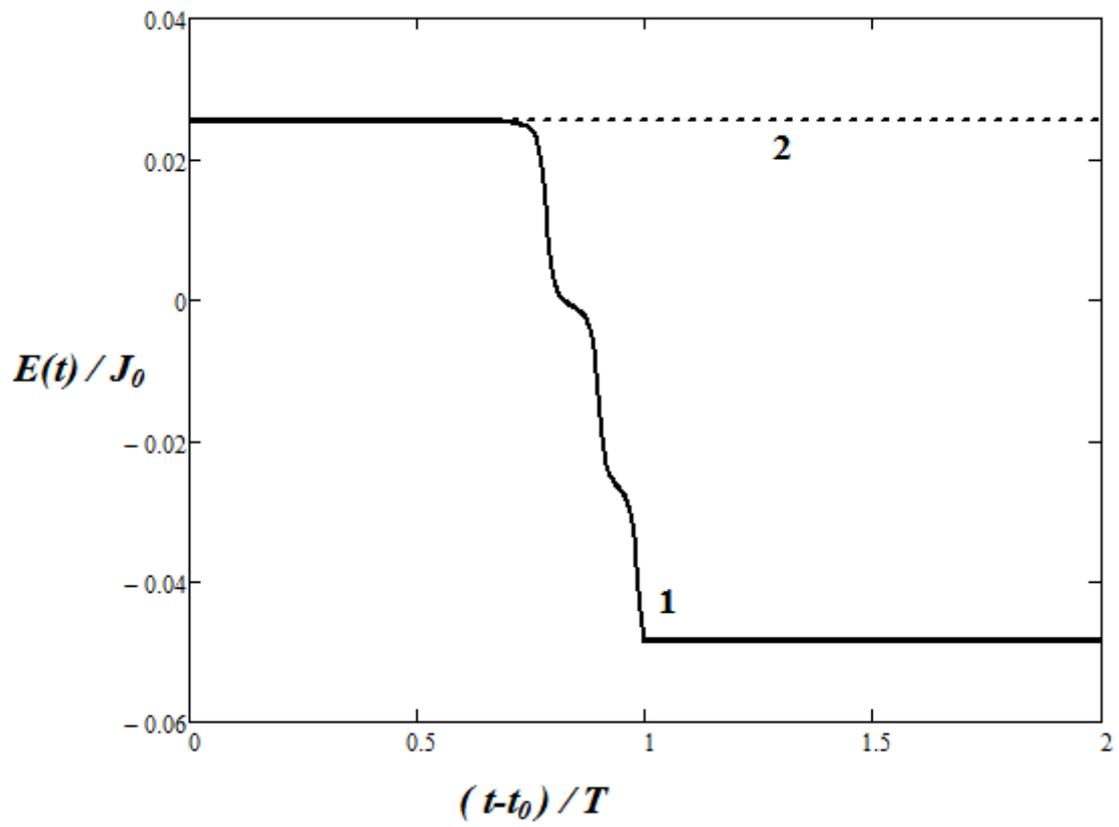

**Fig.3.** The total energy $E(t)$ (in units $J_0 = 5000 \cdot m \cdot r_0^2 / T^2$) versus time $t \geq t_0$. Curves 1 and 2 were calculated respectively for parameters of fig.1,*a* and fig.1,*b*.